\documentstyle[preprint,revtex]{aps}
\begin{document}
\draft
\preprint{1}
\begin{title}
Alpha scattering and capture reactions in the A = 7 system
at low energies
\end{title}
\author{P. Mohr, H. Abele, R. Zwiebel, G. Staudt}
\begin{instit}
Physikalisches Institut der Universit\"at T\"ubingen, D-7400 T\"ubingen,
Germany
\end{instit}
\author{H. Krauss, H. Oberhummer}
\begin{instit}
Institut f\"ur Kernphysik, TU Wien, A-1040 Wien, Austria
\end{instit}
\author{A. Denker, J.W. Hammer, G. Wolf}
\begin{instit}
Institut f\"ur Strahlenphysik, Univ. Stuttgart, D-7000 Stuttgart,
Germany
\end{instit}
\begin{abstract}
Differential cross sections for $^3$He-$\alpha$ scattering were
measured in the energy range up to 3 MeV. These data together with
other available experimental results for $^3$He $+ \alpha$ and $^3$H
$+ \alpha$ scattering were analyzed in the framework of the optical
model using double-folded potentials. The optical potentials obtained
were used to calculate the astrophysical S-factors of the capture
reactions $^3$He$(\alpha,\gamma)^7$Be and $^3$H$(\alpha,\gamma)^7$Li,
and the branching ratios for the transitions into the two final $^7$Be
and $^7$Li bound states, respectively. For
$^3$He$(\alpha,\gamma)^7$Be excellent agreement between calculated
and experimental data is obtained. For $^3$H$(\alpha,\gamma)^7$Li a
$S(0)$ value has been found which is a factor of about 1.5 larger
than the adopted value. For both capture reactions a similar
branching ratio of $R = \sigma(\gamma_1)/\sigma(\gamma_0) \approx 0.43$
has been obtained.
\end{abstract}

\narrowtext
\section{INTRODUCTION}
\label{sec:intro}
The reaction $^3$He$(\alpha,\gamma)^7$Be determines together with
other reactions the branching ratio between the ppI and (ppII +
ppIII) chain in hydrogen burning of main-sequence stars. The magnitude
of the cross section of this capture reaction is of special interest
for the solar neutrino problem \cite{bahc89}. The mirror reaction
$^3$H$(\alpha,\gamma)^7$Li is the main source for $^7$Li production
in primordial nucleosynthesis \cite{stei85}.

Experimental data for the $^3$He$(\alpha,\gamma)^7$Be cross section
at subCoulomb energies have first been obtained by Parker and
Kavanagh \cite{park63}. Further experiments observing
the capture $\gamma$-rays have been performed
by Nagatani et al. \cite{naga69}, Kr\"awinkel et al. \cite{krae82},
Osborne et al. \cite{osbo82}, Alexander et al. \cite{alex84}, and
Hilgemeier et al. \cite{hilg88}. Capture cross sections observing the
decay of the $^7$Be residual nucleus have been measured by Osborne et
al. \cite{osbo82}, Robertson et al. \cite{robe83}, Volk et al.
\cite{volk83} and Hilgemeier et al. \cite{hilg88}. Theoretically,
Tombrello and Parker \cite{tomb63} have first succeeded in describing
the energy dependence of the astrophysical S-factor very well in a
direct-capture model. Further calculations in the framework of the
potential model have been carried out by Kim et al. \cite{kim81} and
Buck et al. \cite{buck85}. Analyses using microscopic theories based
on the resonating group method have been performed by Walliser et al.
\cite{wall84}, Kajino and Arima \cite{kaji84}, Mertelmeier and
Hofmann \cite{mert86}, Langanke \cite{lang86}, Kajino \cite{kaji86}
and Liu et al. \cite{liu86}. The adopted value for the astrophysical
S-factor chosen by Bahcall \cite{bahc89} to calculate the expected
solar neutrino flux is $S(0) = (0.54 \pm 0.03)$ keV b.

In the case of the $^3$H$(\alpha,\gamma)^7$Li reaction three sets of
experimental data at subCoulomb energies are  published by
Griffith et al. \cite{grif61}, Schr\"oder et al. \cite{schr87}, and
Burzynski et al. \cite{burz87}.
Several microscopic calculations have been
performed in order to analyze this reaction
\cite{kaji84,mert86,lang86,kaji86,kaji85,altm88}. The presently
adopted S-factor value used in standard hot big bang model studies is
given by $S(0) = 0.064$ keV b \cite{fowl67}.

In this work we analyze the experimental data of both capture
reactions, $^3$He$(\alpha,\gamma)^7$Be and
$^3$H$(\alpha,\gamma)^7$Li, for energies $E_{\rm CM} \leq 1.4$ MeV and 0.6
MeV, respectively. The calculations were performed in the framework
of the direct-capture model. The most important ingredients in this
model are the optical potentials for the bound and scattering states.
These potentials are determined using the folding procedure. The
strengths of these potentials are adjusted to the experimental
scattering data. Therefore, we measured the differential cross
sections for $^3$He -- $^4$He elastic scattering in the range
$E_{\rm lab}~(^3$He) $\leq 3$ MeV.

In the next section we describe the direct capture model and the
folding procedure for the optical and bound state potentials. In
Sec.~\ref{sec:clust} we present the $^3$He -- $^4$He and $^3$H -- $^4$He
cluster
potentials derived from elastic scattering measurements. Finally, in
Sec.~\ref{sec:capreac} the results for the astrophysical S-factors of the
capture
reactions are given and compared with the experimental data. A
summary is given in Sec.~\ref{sec:summ}.

\section{DIRECT CAPTURE MODEL AND FOLDING PROCEDURE}
\label{sec:dircap}
Potential models are based on the description of the dynamics of
nuclear processes by a Schr\"odinger equation with local potentials
in the entrance and  exit channels. Such models are the Optical
Model (OM) for elastic scattering, the Distorted-Wave Born
Approximation (DWBA) for transfer and the Direct Capture Model
(DC) for direct capture reactions.

The most important ingredients in the potential models are the wave
functions for the scattering and bound states in the entrance and exit
channels. In calculations performed by our group the potentials are determined
by using the folding procedure. In this approach the number of
open parameters is reduced considerably compared to more phenomenological
potentials
(e.g.~Saxon--Woods potentials). The nuclear densities
are derived from nuclear charge distributions \cite{devr87} and folded
with an energy
and density dependent nucleon--nucleon (NN) interaction
$v_{\rm eff}$ \cite{kobo84,ober91,abel93}:
\begin{equation}
V(R) = \lambda V_{\rm F}(R) = \lambda \int \int
\rho_{a} (\vec{r_{1}}) \rho_{A} (\vec{r_{2}}) v_{\rm eff}(E,\rho_a,\rho_A,s)
d\vec{r_{1}} d\vec{r_{2}} \quad .
\end{equation}
The variable $s$ in the NN interaction term is given by
\begin{equation}
s = |\vec{R} + \vec{r_{2}} - \vec{r_{1}}|
\end{equation}
with $\vec{R}$ being the separation of the centers of mass of the two colliding
nuclei. The normalization factor $\lambda$ accounts for Pauli
repulsion effects and dispersive parts in the potential $V(R)$ which are
not included in the folding potential $V_{\rm F}(R)$. This parameter
can be adjusted to elastic scattering data and/or to bound and resonant
state energies of nuclear cluster states. At the low energies
considered in the nucleosynthesis often the imaginary term in the
potential can be neglected. Therefore, in the potential model
combined with the folding procedure for the potential the reaction
cross sections can be calculated in many cases without any free
parameter.

The DC cross section is given by \cite{kim87}
\begin{eqnarray}
\sigma^{{\rm DC}} & = &
\int d\Omega \:{d \sigma^{{\rm DC}} \over d \Omega} \nonumber \\
& = &
\int d\Omega \: 2 \left( e^2 \over \hbar c \right) \left( \mu c^2 \over \hbar c
\right)
\left( k_\gamma \over k_a \right)^{\!\! 3} {1 \over 2 I_A + 1} \;
{1 \over 2 S_a + 1}
\sum_{M_A M_a M_B \sigma}
\mid T_{M_A M_a M_B, \sigma} \mid^2~~ .
\end{eqnarray}
The quantities $I_{A}$, $I_{B}$ and $S_{a}$
($M_{A}$, $M_{B}$ and $M_{a}$)
are the spins (magnetic quantum numbers) of the target nucleus $A$,
residual nucleus $B$ and projectile $a$, respectively.
The reduced mass in the entrance channel is given by $\mu$. The polarisation
$\sigma$ of the electromagnetic radiation can be $\pm 1$. The wave
number in the entrance channel and for the emitted radiation
is given by $k_{a}$ and $k_{\gamma}$, respectively.

The multipole expansion of the transition matrices $T_{M_A M_a M_B, \sigma}$
including electric dipole (E1) and quadrupole (E2)
transitions as well as magnetic dipole (M1) transitions is given by
\begin{equation}
T_{M_A M_a M_B, \sigma}=
T_{M_A M_a M_B, \sigma}^{\rm E1}
\, d_{\delta \sigma}^1(\theta) + T_{M_A M_a M_B, \sigma}^{\rm E2}
\, d_{\delta \sigma}^2(\theta) + T_{M_A M_a M_B, \sigma}^{\rm M1}
\, d_{\delta \sigma}^1(\theta) \quad .
\end{equation}
The rotation matrices depend on the angle between
$\vec{k}_a$ and $\vec{k}_\gamma$ which is denoted by $\theta$,
where $\delta = M_A + M_a - M_B$.

Defining
\begin{eqnarray}
C({\rm E}1) & = &
i\, \mu\, \left( {Z_a \over m_a} - {Z_A \over m_A} \right) \quad ,\label{3}\\
	\rule[0mm]{0cm}{8mm}
C({\rm E}2) & = &
{k_\gamma \over \sqrt{12}}\, \mu^2
\left( {Z_a \over m_a^2} + {Z_A \over m_A^2} \right) \quad , \label{4}
\end{eqnarray}
we can write for the transition matrices for the electric
dipole (E${\cal L}$ = E1) or quadrupole (E${\cal L}$ = E2) transition
\begin{eqnarray}
T_{M_A M_a M_B, \sigma}^{{\rm E}{\cal L}}
 &= &
	\sum_{l_a j_a}
	i^{l_a} (l_a\, 0\, S_a\, M_a \mid j_a\, M_a)
	(j_b\, M_B\! -\! M_A\, I_A\, M_A \mid I_B\, M_B) \nonumber\\
&\times &
	 ({\cal L}\, \delta\, j_b\, M_B\! -\! M_A \mid j_a\, M_a)
	\, C({\rm E}{\cal L})
	\, \hat{l}_a\, \hat{l}_b\, \hat{j}_b \nonumber\\
	\rule[0mm]{0cm}{8.3mm}
&\times &
	\, (l_b\, 0\, {\cal L}\, 0 \mid l_a\, 0)
	\: {\cal W}({\cal L}\, l_b\, j_a\, S_a;\, l_a\, j_b)
	\: I_{l_b j_b I_B; l_a j_a}^{{\rm E}{\cal L}}~~.
\end{eqnarray}
In the above expressions $Z_{a}$, $Z_{A}$ and
$m_{a}$, $m_{A}$ are the charge and mass numbers
of the projectile $a$ and target nucleus $A$, respectively. The
quantum numbers for the channel spin in the entrance channel
and for the transferred angular momentum are denoted by $j_{a}$ and
$j_{b}$, respectively.

For magnetic dipole transitions (M${\cal L}$ = M1) we obtain
\begin{eqnarray}
T_{M_A M_a M_B, \sigma}^{{\rm M}{\cal L}}
&= &
	\sum_{l_a j_a}
	i^{l_a}\, \sigma \, \Biggl\{ (l_a\, 0\, S_a \, M_a \mid j_a\, M_a)
	(j_b\, M_B\! -\! M_A\, I_A\, M_A \mid I_B\, M_B) \nonumber\\
	\rule[-4mm]{0cm}{9mm}
&\times &
	(1\, \delta \, j_b\, M_B\! -\! M_A \mid j_a\, M_a) \nonumber\\
&\times &
	\Biggl[ \mu \Biggl( {Z_A \over m_A^2} + {Z_a \over m_a^2}
	\Biggr) \, \hat{l}_b\,
	\hat{j}_b\, \sqrt{\, l_a(l_a+1)}
	 \: {\cal W}(1\, l_a\, j_a\, S_a;\, l_a\, j_b) \nonumber \\
&+ &
	2 \mu_a (-1)^{j_b-j_a}\, \hat{S}_a\, \hat{j}_b\,
	\sqrt{\, S_a(S_a+1)}\:
	{\cal W}(1\, S_a\, j_a\, l_a;\, S_a\, j_b) \Biggr] \nonumber\\
	\rule[-4mm]{0cm}{9mm}
&- &
	(l_a\, 0\, S_a\, M_a \mid j_a\, M_a)
	(j_a\, M_a\, I_A\, M_B\! -\! M_a \mid I_B\, M_B) \nonumber\\
	\rule[-4mm]{0cm}{9mm}
&\times &
	(I_A\, M_B\! -\! M_a\, 1\, \delta\, \mid I_A\, M_A) \nonumber \\
&\times &
	\mu_A \, \delta_{j_a j_b}
	\sqrt{\, (I_A+1)/I_A} \Biggr\}
	\Biggl\{ {\hbar c \over 2 m_{\rm p} c^2}
	 \Biggr\}\, \delta_{l_a l_b}
	\: \hat{l}_a\: I_{l_b j_b I_B; l_a j_a}^{\rm M1}
	~~,\label{kim4}
\end{eqnarray}
where {\cal W} is the Racah coefficient, the $\mu_i$ are the magnetic moments
and
$m_{\rm p}$ is the mass of the proton.

The overlap integrals in Eqs.~(7) and (8) are given as
\begin{equation}
I_{l_b j_b I_B; l_a j_a}^{{\rm E}{\cal L}} = \int dr\: u_{NLJ}(r) \:
{\cal O}^{{\rm E}{\cal L}}(r) \: \chi_{l_a j_a} (r)
\end{equation}
for the electric
dipole (E${\cal L}$ = E1) or quadrupole (E${\cal L}$ = E2)
transition, and by
\begin{equation}
I_{l_b j_b I_B; l_a j_a}^{\rm M1} = \int dr\: u_{NLJ}(r) \:
{\cal O}^{\rm M1}(r) \: \chi_{l_a j_a} (r)
\end{equation}
for the magnetic dipole transition (M${\cal L}$ = M1).

The radial part of the bound state wave function
in the exit channel and the scattering wave function
in the entrance channel is given by $u_{NLJ}(r)$
and $\chi_{l_a j_a} (r)$, respectively.
The radial parts of the electromagnetic multipole operators
are \cite{bail67}
\begin{eqnarray}
{\cal O}^{\rm M1}(r)&=&
{1\over 2\rho}\left[
\sin \rho + \rho \cos \rho\right] \quad ,\label{9}\\
{\cal O}^{\rm E1}(r)&=&
{3\over \rho^3}\left[ (\rho^2 - 2) \sin \rho +
2 \rho \cos \rho \right] r \quad ,\label{10}\\
{\cal O}^{\rm E2}(r)&=&
{15 \over \rho^5}\left[ (5 \rho^2 - 12)
\sin \rho + (12 - \rho^2) \rho \cos \rho \right] r^2 \quad .\label{11}
\end{eqnarray}
In the long wave-length approximation -- applicable
in our case, since $\rho=k_\gamma r \ll 1$ -- these
quantities reduce to
\begin{eqnarray}
{\cal O}^{\rm M1}(r)&\simeq & 1 \quad ,\label{12}\\
{\cal O}^{\rm E1}(r)&\simeq & r \quad ,\label{13}\\
{\cal O}^{\rm E2}(r)&\simeq & r^2 \quad .\label{14}
\end{eqnarray}
\bigskip
\section{$^3$H\protect\lowercase{e} -- $^4$H\protect\lowercase{e} and
$^3$H --
$^4$H\protect\lowercase{e} CLUSTER POTENTIALS}
\label{sec:clust}

At incident energies $E_{\rm lab}(^3$He) $\approx 3 - 10$ MeV
differential cross sections for the elastic $^3$He -- $^4$He
scattering are well known \cite{barn64,spig67,chua71,boyk72,hard72}.
In order to obtain cross section data at very low energies we have
measured these scattering processes at 20 energies in the range from
1 to 3.3 MeV using the windowless, differential pumped and
recirculating gas target system RHINOCEROS installed at the Stuttgart
Dynamitron accelerator \cite{hamm}. In our experiment we used the jet
configuration in which a supersonic jet produced by a laval nozzle
serves as a nearly pointlike target zone with high
density~\cite{knee87}. The small
and fixed size of this zone allows a good determination of angular
distributions.

The target zone is centered in a 50 cm-diameter scattering chamber.
The detector system consisted of ten surface-barrier detectors
mounted at fixed positions. $^3$He-beam intensities between 5 and 30
$\mu$A were used. In order to normalize the data (assuming that the
$^3$He--$^{20}$Ne scattering is dominated in this energy range by
Rutherford interaction), a small quantity of $^{20}$Ne gas was
admixed to the  $^4$He  gas in the jet.

The experimental differential cross sections for the
$^3$He -- $^4$He scattering are shown together with older data of
Barnard et al. \cite{barn64} and Chuang \cite{chua71} in
Fig.~\ref{expdif}. The
results of different phase-shift analyses
\cite{spig67,boyk72,hard72} for projectile energies
$E_{\rm lab}(^3$He) $>$ 3 MeV are presented in Fig.~\ref{phase}.

We have calculated the differential cross sections and phase shifts in
the framework of the OM. For the calculation of the real part
of the optical $^3$He --
$^4$He potential we used the folding procedure as described in
Sec.~\ref{sec:dircap}. The folding potentials (Eq. 1) were determined using
the computer code DFOLD \cite{abel}. The imaginary part was neglected because
the flux into
other channels is very small. Together with the spin-orbit term, the
optical potential is given by
\begin{equation}
V(R) = \lambda V_{\rm F}(R) + \lambda_{\rm so} {1 \over R} {dV_F(R) \over dR}
\vec{L} \cdot \vec{s}
\end{equation}
with a spin-orbit normalization factor $\lambda_{\rm so}$.

As result of a fit to the experimental data given in
Figs.~\ref{expdif} and \ref{phase}
we obtain a parity-dependent potential. The normalization factors
$\lambda$ together with the volume integrals per nucleon pair $J_{\rm
R}$ are
listed in the upper part of Table~\ref{normaliz}. The spin-orbit normalization
factor $\lambda_{\rm so} = -0.162$ fm$^2$ has been determined from the
splitting of the phase shifts for the $L = 3$ doublet. The agreement
between the experimental data and the results of the OM calculation
is excellent in the whole energy  range up to 14 MeV as can be seen in
Figs.~\ref{expdif} and \ref{phase}.

For the $^3$H -- $^4$He system phase-shift analyses of experimental
cross section data have been performed in the energy range
$E_{\rm lab}(^3$H) $= 3 - 10$ MeV \cite{spig67}. We have calculated
these phase-shifts in the OM. The optical potential is again
determined using the folding procedure. The result of our OM fit to
the phase--shift data is shown in Fig.~\ref{phases}. The normalization factors
$\lambda$ together with the volume integrals per nucleon pair $J_{\rm
R}$ are
presented in the lower part of Table~\ref{normaliz}. For the spin-orbit
normalization factor $\lambda_{\rm so}$ now we obtain $\lambda_{\rm so} =
-0.136$ fm$^2$. Again the agreement between the experimental and
calculated data is satisfactory (s. Fig.~\ref{phases}).

The volume integrals of the parity-dependent potentials for
$\alpha$-scattering on the mirror nuclei $^3$He and $^3$H only differ
by about 0.5\% and 3\% for the even and odd partial waves,
respectively. The values of the volume integrals for the even partial
waves in $^3$He -- $\alpha$ and $^3$H -- $\alpha$ scattering is
comparable with the value $J_{\rm R} = 445.7$ MeV fm$^3$ obtained in the
analysis of $\alpha$ -- $\alpha$ scattering using the folding
procedure \cite{abel92}. Furthermore, the values of $J_{\rm R}$ for $^3$He
-- $\alpha$ scattering are compatible with the results of a
systematic analysis of $^3$He-scattering on several nuclei
\cite{goer79,tros80}.

In a next step we used the double-folded potential as a suitable
cluster--cluster potential and calculated bound states and
single-particle (single-cluster) resonances. The wave function
$u_{NLJ}(r)$ which describes the relative motion of the respective
$^3$He -- $\alpha$ and $^3$H -- $\alpha$ system is characterized by
the node number $N$ and the orbital angular momentum $L$. The $N$ and
$L$ values are related to the corresponding quantum numbers $n_i$ and
$l_i$ of the three nucleons forming the $^3$He and $^3$H cluster,
respectively:
\begin{equation}
Q = 2N + L = \sum_{i=1}^3 2 n_i + l_i = \sum_{i=1}^3 q_i = 3~~.
\end{equation}
Thus for both systems $^7$Be = $^3$He $\otimes~\alpha$ and $^7$Li $=~^3$H
$\otimes~\alpha$ one expects two cluster states with $L = 1~~(N =
1)$ and $L = 3~~(N = 0)$. Both states split into doublets with $J = L
\pm 1/2$ because of
the spin-orbit potential resulting from the motion of the $A = 3$
particle with spin 1/2 in the field of the $\alpha$-particle.

In the calculations of the cluster states
the centroid energies of both the bound $(L = 1 : J^\pi = 3/2^-,1/2^-)$
and quasi-bound $(L = 3 : J^\pi = 7/2^-,5/2^-)$ states for $^7$Be and
$^7$Li (s. Fig.~\ref{schem}) are reproduced by the central parts of the $^3$He
-- $\alpha$ and $^3$H -- $\alpha$ odd potential, respectively. The
splitting of the energies of the quasi-bound state doublets $(L = 3)$
in $^7$Be and $^7$Li is reproduced by the spin-orbit potential as
determined by the OM calculation. However, for the bound-state
doublets $(L = 1)$ in $^7$Be and $^7$Li smaller spin-orbit potentials
are necessary. For the energy splitting of the bound-state doublets a
spin-orbit normalization factor $\lambda_{\rm so} = -0.07$ fm$^2$
has to be used for both nuclei $^7$Be and $^7$Li.

As a further test of our folding potential we calculated the charge
distribution of the $^7$Li nucleus (in its ground state) by folding
the experimental charge distribution of $^3$H and $^4$He, which we
have already used in our double-folding procedure, with the radial
wave function $u_{N=1,L=1,J=3/2}$ and  by assuming a spherical shape for the
folded
distribution. The result of this calculation is shown in
Fig.~\ref{dichte}
together with the experimental charge distribution, as measured by
electron scattering on $^7$Li.
It can be seen that the calculation overestimates the
experimental density in the nuclear interior, whereas in the surface
region, near $r = 2.5$ fm, the experimental values are slightly
underestimated. But the rms radii of both distributions are almost
identical: $< r^2 >^{1/2} \approx 2.40$ fm.
\bigskip
\section{CAPTURE REACTIONS}
\label{sec:capreac}
In Fig.~\ref{schem} a schematic presentation of the direct capture processes
$^3$He$(\alpha,\gamma)^7$Be and $^3$H$(\alpha,\gamma)^7$Li is given.
In the low energy range capture transitions can only occur into the
ground and first excited states of $^7$Be and $^7$Li, respectively.

The theoretical cross section $\sigma^{\rm th}$ is obtained from the DC
cross section  $\sigma^{\rm DC}$, given in Eq. 3, as sum over both final
states $i = 1,2$ by
\begin{equation}
\sigma^{\rm th} = \sum_i C^2_iS_i\sigma^{\rm DC}_i~~.
\end{equation}
The computation of the cross section $\sigma^{\rm th}$ was performed
using the computer code TEDCA~\cite{krau}. As input three data sets
are necessary: (i) isospin Clebsch-Gordan coefficients, (ii)
spectroscopic factors $S_i$ which specify the cluster probability of
the final states $^7$Be $=~^3$He $\otimes~\alpha$ and $^7$Li $=~^3$H
$\otimes~\alpha$, respectively, and (iii) optical potentials for the
calculation of the wave functions in the entrance and exit channel.
In our case the Clebsch-Gordan coefficients are $C_i = 1$. The
spectroscopic factors $S_i$ have been taken from the work of Kurath
and Millener~\cite{kura75}. The numerical values are given as $S_1 =
1.174$ and $S_2 = 1.175$. In order to calculate the bound state wave
function in the exit channel and the scattering wave function in the
entrance channel (Eqs. 9,10), the folded potentials are used which
have already been determined in Secs.~\ref{sec:dircap} and
\ref{sec:clust}. That means that all
the necessary information for the calculation of the DC reaction is
known and no parameter has to be adjusted to the experimental capture
reaction data.

In Fig.~\ref{calc} the experimental values of the astrophysical
$S$-factor for the reactions
$^3$He$(\alpha,\gamma)^7$Be \cite{krae82,osbo82,hilg88} and
$^3$H$(\alpha,\gamma)^7$Li \cite{grif61,schr87} are shown together
with the results of DC calculations using parity-dependent folding
potentials. The experimental data of Kr\"awinkel et~al. \cite{krae82}
are renormalized by a factor of 1.4 as suggested by Hilgemeier et~al.
\cite{hilg88}.

For the $^3$He$(\alpha,\gamma)^7$Be reaction the agreement between
the experimental and calculated data is excellent.
A linear extrapolation for $E \to 0$ gives $S(0) = 0.516$ keV
b and $\dot{S}(0) = -3.67 \cdot 10^{-4}$ b ($E$ in keV).
The values of $S(0)$ agree excellently with the
experimentally determined $S(0)$ factor, for which~\cite{hilg88} gives
a weighted average of $S(0) = (0.51 \pm 0.02)$ keV   b, and with
the adopted value~\cite{bahc89} of $S(0) = (0.54 \pm 0.03)$ keV
  b. The calculated branching ratio $R$ for the transitions to
the first excited state and the ground state has the value $R =
0.43$. This value was found to be nearly energy-independent in the
energy range 0 -- 1.4 MeV and agrees well with both the experimental
data~\cite{krae82,osbo82,hilg88} and the results of microscopic
calculations~\cite{altm88}.

The two presently available measurements of the low-energy
$^3$H$(\alpha,\gamma)^7$Li reaction differ from each other by roughly
30\% in total magnitude as well as in their determination of the
branching ratios for the transitions into the two final bound states.
The calculation within the potential model gives results for the
absolute magnitude of the S-factor which favour rather the older
experimental data of Griffith et~al. \cite{grif61}. The calculated
branching ratio is again nearly energy-independent and has
likewise a value of
$R = 0.43$. This branching ratio, however, is noticeably larger
than the experimentally observed value $R = 0.32 \pm 0.01$
\cite{schr87}. Our value agrees with the data of \cite{grif61} which
give an energy-independent average of $R = 0.43$ \cite{altm88}. The
above considerations are comparable with the results of microscopic RGM
calculations \cite{altm88}.
A linear extrapolation for $E \to 0$ gives $S(0) = 0.100 $ keV
b and $\dot{S}(0) = -1.02  \cdot 10^{-4}$ b ($E$ in keV).
The values of $S(0)$ are nearly twice as large as the
adopted value of 0.064 keV b \cite{fowl67}, obtained from an
energy-independent extrapolation, and somewhat smaller than the
extrapolated value 0.14 keV  b extracted from the data of
Schr\"oder et~al. \cite{schr87}. However, it is in excellent agreement with the
results of different RGM calculations
\cite{mert86,lang86,kaji86,kaji85,altm88} giving a value of $S(0)
\approx 0.1$ keV b.

In Fig.~\ref{mult} the multipole contributions for both capture reactions are
shown. The main contribution is the E1 transition. Because of the
missing centrifugal barrier, the DC transitions from the s-wave in
the entrance channel to the final $L = 1$ states are dominating. For the
$^3$H$(\alpha,\gamma)^7$Li reaction the contributions of the higher
partial waves for the total S-factor can even be neglected. As can be
seen on the right-hand side of Fig.~\ref{mult} the curve of the total S-factor
is almost identical to the s-wave contribution. The Coulomb barrier is lower
for $^3$H$(\alpha,\gamma)^7$Li than for the mirror reaction.
Therefore, the influence of the centrifugal barrier is more
pronounced.

As already discussed in Sec.~\ref{sec:clust}, a parity--dependent
optical potential is necessary to describe the experimental
scattering data. We calculated the astrophysical S-factor with a
parity--independent potential using the $\lambda$-parameter for the
dominating even partial waves given in Table~\ref{normaliz}. As shown in
Fig.~\ref{fact},
it is impossible to reproduce the experimental data with such a
parity-independent potential. The enhancement for energies $E \geq
0.7$ MeV is due to the M1-contribution of the p-wave.

We also tested the sensitivity of the astrophysical S-factor by
changing the strength of the optical potentials in the entrance
channel. A variation of the potential depth of $\pm 1\%$ leads to an
energy-independent  change of the S-factor of $\pm 2\%$. The S-factor
increases with growing potential depth. The reason for this behaviour
is that with growing nuclear potential depth the Coulomb barrier
becomes smaller and therefore the penetration probability is
enhanced.
\newpage
\section{SUMMARY}
\label{sec:summ}
Differential cross sections for the elastic scattering of $^3$He
particles on $^4$He and phase shifts of both $^3$He -- $^4$He and
$^3$H -- $^4$He scattering have been analyzed up to energies of about
10 MeV in the framework of the optical model. The potential was
deduced by a double-folding procedure using a density-dependent
effective nucleon-nucleon interaction. The experimental data are
described satisfactorily by this optical-model calculation.

Using the double-folded $^3$He - $\alpha$ and $^3$H - $\alpha$
potentials as cluster-cluster potentials, we calculate the bound and
quasi-bound doublet states in $^7$Be and $^7$Li, respectively. The
excitation energies of these states as well as the charge
distribution of $^7$Li are well reproduced in this potential model.

The optical potentials
obtained from the fit to the elastic scattering data have been used
to calculate the astrophysical S-factors of
$^3$He$(\alpha,\gamma)^7$Be and $^3$H$(\alpha,\gamma)^7$Li within the
direct capture model. Using this method no parameter has to be
adjusted to the experimental reaction data.

In the case of the reaction $^3$He$(\alpha,\gamma)^7$Be we obtain
$S(0) = 0.516$ keV b. This value is in excellent agreement
with both the average of the experimental data~\cite{hilg88} and the
adopted value~\cite{bahc89}. The branching ratio for the transitions
to the first excited state and the ground state results in $R =
0.43$, likewise in very good agreement with the known experimental
and theoretical data.

For the reaction $^3$H$(\alpha,\gamma)^7$Li the three presently
available measurements differ from each other by roughly 30\% in
total magnitude as well as in their determination of the branching
ratio for the transitions into the two final $^7$Li bound states. In
agreement with calculations which have been done in the framework of
the Resonating Group method or on the basis of a microscopic
potential model, our calculations predict the $S$-factor to increase
with decreasing energy resulting in $S(0) = 0.10$ keV b. This
value is a factor of about 1.5 larger than the adopted
value~\cite{fowl67}. For the branching ratio we obtain $R = 0.43$.
This value is in good agreement with some theoretical results, but is
not compatible with the recently measured value of $R = 0.32 \pm
0.01$~\cite{schr87}.

\section{ACKNOWLEDGEMENT}
\label{sec:ack}
We like to thank the Deutsche Forschungsgemeinschaft (DFG-project
Sta/2-1), the \"Osterreichische Nationalbank (project 3924), and
the Fonds zur F\"orderung der wissenschaftlichen Forschung in
\"Osterreich (project P 8806--PNY).
\newpage
%\mediumtext
\begin{table}
\caption{Normalization factors $\lambda$ and volume integrals per
nucleon $J_{\rm R}$ of the optical potentials}
\label{normaliz}
\begin{tabular}{crcc}
&\multicolumn{1}{c}{partial wave}
&\multicolumn{1}{c}{$\lambda$}
&\multicolumn{1}{c}{$J_{\rm R}$(MeV/fm$^3$)} \\
\tableline
\multicolumn{1}{c}{$^3$He -- $\alpha$}
&\multicolumn{1}{r}{even (s,d)}
&\multicolumn{1}{c}{1.452}
&\multicolumn{1}{c}{469.0} \\
&\multicolumn{1}{r}{odd (p,f)}
&\multicolumn{1}{c}{1.844}
&\multicolumn{1}{c}{595.6} \\
\\
\multicolumn{1}{c}{$^3$H -- $\alpha$}
&\multicolumn{1}{r}{even (s,d)}
&\multicolumn{1}{c}{1.525}
&\multicolumn{1}{c}{466.8} \\
&\multicolumn{1}{r}{odd (p,f)}
&\multicolumn{1}{c}{1.890}
&\multicolumn{1}{c}{578.5}\\
\end{tabular}
\end{table}

\newpage

\figure{Experimental differential cross section for the elastic
$^3$He -- $^4$He scattering for projectile energies between
$E_{\rm lab}(^3$He) $= 1.2$ and 3.0 MeV together with data of Barnard et
al. \cite{barn64} and Chuang \cite{chua71}. The solid lines are the
result of the present OM calculation, the dashed lines give
Rutherford scattering. \label{expdif}}

\figure{Phase shifts deduced from experimental $^3$He -- $^4$He
scattering data given by Spiger and Tombrello $(\Delta$,
\cite{spig67}), Boykin et~al. ($\diamond$, \cite{boyk72}) and Hardy
et~al. ($\sqcap$, \cite{hard72}). The solid lines are the result of the
present OM calculation. \label{phase}}

\figure{Phase shifts deduced from experimental $^3$H -- $^4$He
scattering data given by Spiger and Tombrello \cite{spig67}. The
solid lines are the result of the present OM calculations.
\label{phases}}

\figure{Comparison of the experimental charge distribution of
$^7$Li~\cite{devr87} (dashed line) with the distribution calculated in the
potential model (solid line). \label{dichte}}

\figure{Schematic presentation of the level scheme for the reactions
$^3$He$(\alpha,\gamma)^7$Be and $^3$H$(\alpha,\gamma)^7$Li.
\label{schem}}

\figure{Calculated astrophysical S-factor using the potential model
compared with the experimental data for $^3$He$(\alpha,\gamma)^7$Be
(upper part: closed circles~\cite{krae82}, open circles~\cite{osbo82},
triangles~\cite{hilg88}), and
$^3$H$(\alpha,\gamma)^7$Li (lower part: open circles~\cite{grif61}, closed
circles~\cite{schr87}, triangles~\cite{burz87}). \label{calc}}

\figure{Multipole contributions to the astrophysical S-factors for
the reaction $^3$He$(\alpha,\gamma)^7$Be (left-hand side) and
$^3$H$(\alpha,\gamma)^7$Li (right-hand side). The experimental data
are the same as in Fig.~\ref{calc}. \label{mult}}

\figure{S-factor for $^3$He$(\alpha,\gamma)^7$Be calculated with a
parity-independent potential in the entrance channel. The
experimental data are the same as in Fig.~\ref{calc}. \label{fact}}


\newpage
\begin{references}
\bibitem{bahc89} J.N. Bahcall, Neutrino Astrophysics, Cambridge
University Press, Cambridge 1989
\bibitem{stei85} G. Steigman, Nucleo-synthesis-challenges and new
developments, in: W.D. Arnett, J.W. Truran (eds.) p. 48, Chicago
Press, Chicago 1985
\bibitem{park63} P.D. Parker and R.W. Kavanagh, Phys. Rev. {\bf
131},2578 (1963)
\bibitem{naga69} K. Nagatani, M.R. Dwarakanath, and D. Ashery, Nucl.
Phys. {\bf A128}, 325 (1969)
\bibitem{krae82} H. Kr\"awinkel, H.W. Becker, L. Buchmann, J.
G\"orres, K.U. Kettner, W.E. Kieser, R. Santo, P. Schmalbrock, H.P.
Trautvetter, A. Vlieks, C. Rolfs, J.W. Hammer, R.E. Azuma, and W.S.
Rodney,  Z. Phys. {\bf A304}, 307 (1982)
\bibitem{osbo82} J.L. Osborne, C.A. Barnes, R.W. Kavanagh, R.M.
Kremer, G.J. Mathews, J.L. Zyskind, P.D. Parker, and A.J. Howard,
Phys. Rev. Lett. {\bf 48}, 1664 (1982); Nucl. Phys. {\bf A419}, 115
(1984)
\bibitem{alex84} T.K. Alexander, G.C. Ball, W.N. Lennard, H.Geissel,
and H.-B. Mak, Nucl. Phys. {\bf A427}, 526
(1984)
\bibitem{hilg88} M. Hilgemeier, H.W. Becker, C. Rolfs, H.P.
Trautvetter, and J.W. Hammer, Z. Phys. {\bf A329}, 243
(1988)
\bibitem{robe83} H.G.H. Robertson, P. Dyer, T.J. Bowles, R.E. Brown,
N. Jarmie, C.J. Maggiore, and S.M. Austin, Phys. Rev. {\bf C27}, 11
(1983)
\bibitem{volk83} H. Volk, H. Kr\"awinkel, R. Santo, and L. Wallek, Z.
Phys. {\bf A310}, 91 (1983)
\bibitem{tomb63} T.A. Tombrello and P.D. Parker, Phys. Rev. {\bf 131}, 2582
(1963)
\bibitem{kim81} B.T. Kim, T. Izumoto, and K. Nagatani, Phys. Rev. {\bf C23}, 33
(1981)
\bibitem{buck85} B. Buck, R.A. Baldock, and J.A. Rubio,
J. Phys. {\bf G11}, L11 (1985)
\bibitem{wall84} H. Walliser, H. Kanada, and Y.C. Tang,
Nucl. Phys. {\bf A419}, 133 (1984)
\bibitem{kaji84} T. Kajino and A. Arima, Phys. Rev. Lett. {\bf 52}, 739
(1984)
\bibitem{mert86} T. Mertelmeier and H.M. Hofmann, Nucl. Phys. {\bf A459}, 387
(1986)
\bibitem{lang86} K. Langanke, Nucl. Phys. {\bf A457}, 351 (1986)
\bibitem{kaji86} T. Kajino, Nucl. Phys. {\bf A460}, 559 (1986)
\bibitem{liu86} Q.K.K. Liu, H. Kanada, and Y.F. Tang,
Phys. Rev. {\bf 33}, 1561 (1986)
\bibitem{grif61} G.M. Griffith, R.A. Morrow, P.J. Riley, and J.B. Warren,
Can. J. Phys. {\bf 39}, 1397 (1961)
\bibitem{schr87} U. Schr\"oder, A. Redder, C. Rolfs, R.E. Azuma, L.
Buchmann, C. Campbell, J.D. King, and T.R. Donoghue, Phys. Lett. {\bf
B192}, 55 (1987)
\bibitem{burz87} S. Burzynski, K. Czerski, A. Marcinkowski, and P.
Zupranski, Nucl. Phys. {\bf A473}, 179 (1987)
\bibitem{kaji85} T. Kajino, J. Phys. Soc. Jap. {\bf 54}(Suppl), 321
(1985)
\bibitem{altm88} T. Altmeyer, E. Kolbe, T. Warmann, K. Langanke, and H.J.
Assenbaum, Z. Phys. {\bf A330}, 277 (1988)
\bibitem{fowl67} W.A. Fowler, G.R. Caughlan, and B.A. Zimmermann, Ann.
Rev. Astron. Astrophys. {\bf 5}, 525 (1967), {\bf 13}, 69 (1975)
\bibitem{devr87} H. de Vries, C.W. Jager, and C. de Vries, At. Data
and Nucl. Data Tables {\bf 36}, 495 (1987)
\bibitem{kobo84} A.M. Kobos, B.A. Brown, R. Lindsay, and R. Satchler,
Nucl. Phys. {\bf A425}, 205 (1984)
\bibitem{ober91} H. Oberhummer and G. Staudt, Nuclei in the Cosmos,
H. Oberhummer (ed.) Heidelberg 1991, p 29
\bibitem{abel93} H. Abele and G. Staudt, Phys. Rev. {\bf C47}, 742
(1993)
\bibitem{kim87} K.H. Kim, M.H. Park, and B.T. Kim, Phys. Rev. {\bf C35},
363  (1987)
\bibitem{bail67} B.G. Bailey, G.M. Griffiths, and T.W. Donnely, Nucl.
Phys. {\bf A94}, 502 (1967)
\bibitem{barn64} A.C.L. Barnard, C.M. Jones, and G.C. Phillips, Nucl.
Phys. {\bf 50}, 629 (1964)
\bibitem{spig67} R.J. Spiger and T.A. Tombrello, Phys. Rev. {\bf63},
964 (1967)
\bibitem{chua71} L.S. Chuang, Nucl. Phys. {\bf A174}, 399 (1971)
\bibitem{boyk72} W.R. Boykin, S.D. Baker, and D.M. Hardy, Nucl. Phys.
{\bf A195}, 241 (1972)
\bibitem{hard72} D.M. Hardy, R.J. Spiger, S.D. Baker, Y.S.Chen, and
T.A. Tombrello, Nucl. Phys. {\bf A195}, 250 (1972)
\bibitem{hamm} J.W. Hammer, W. Biermayer, T. Griegel, H. Knee and K.
Petkau, Nucl. Inst. Meth., to be published
\bibitem{knee87} H. Knee, diploma thesis, Univ. Stuttgart (1987)
\bibitem{abel} H. Abele, computer code DFOLD, University of
T\"ubingen (unpublished)
\bibitem{abel92} H. Abele, PhD thesis, University of T\"ubingen
(1992)
\bibitem{goer79} R. G\"orgen, F. Hinterberger, R. Jahn, P. von
Rossen, and B. Sch\"uller, Nucl. Phys. {\bf A320}, 296 (1979)
\bibitem{tros80} H.-J. Trost, A. Schwarz, U. Feindt, F.H. Heimlich,
S. Heinzel, J. Hintze, F. K\"orber, R. Lekebusch, P. Lezoch, G.
M\"ock, W. Paul, E. Roick, M. Wolff, J. Worzeck, and U. Strohbusch,
Nucl. Phys. {\bf A337}, 377 (1980)
\bibitem{krau} H. Krauss, computer code TEDCA, TU Wien, (unpublished)
\bibitem{kura75} D. Kurath and D.J. Millener, Nucl. Phys. {\bf A238},
269 (1975)


\end{references}
\end{document}